\documentclass[a4paper,12pt]{article}
\usepackage{graphics}
\usepackage{amsfonts}

\begin{document}
%

\newcommand{\be}{\begin{equation}}
\newcommand{\ee}{\end{equation}}
\newcommand{\bea}{\begin{eqnarray}}
\newcommand{\eea}{\end{eqnarray}}
\newcommand{\bean}{\begin{eqnarray*}}
\newcommand{\eean}{\end{eqnarray*}}
\font\upright=cmu10 scaled\magstep1
\font\sans=cmss12
\newcommand{\ssf}{\sans}
\newcommand{\stroke}{\vrule height8pt width0.4pt depth-0.1pt}
\newcommand{\Z}{\hbox{\upright\rlap{\ssf Z}\kern 2.7pt {\ssf Z}}}
\newcommand{\ZZ}{\Z\hskip -10pt \Z_2}
\newcommand{\C}{{\rlap{\upright\rlap{C}\kern 3.8pt\stroke}\phantom{C}}}
\newcommand{\R}{\hbox{\upright\rlap{I}\kern 1.7pt R}}
\newcommand{\HH}{\hbox{\upright\rlap{I}\kern 1.7pt H}}
\newcommand{\CP}{\hbox{\C{\upright\rlap{I}\kern 1.5pt P}}}
\newcommand{\identity}{{\upright\rlap{1}\kern 2.0pt 1}}
\newcommand{\half}{\frac{1}{2}}
\newcommand{\quart}{\frac{1}{4}}
\newcommand{\pr}{\partial}
\newcommand{\bm}{\boldmath}
\newcommand{\I}{{\cal I}} 
\newcommand{\M}{{\cal M}}
\newcommand{\N}{{\cal N}}
\newcommand{\e}{\varepsilon}

\thispagestyle{empty}
\vskip 3em
\begin{center}
{{\bf \huge Quantum Statistical Mechanics of Vortices
}} 
\\[15mm]

{\bf \large N.~S. Manton\footnote{email: N.S.Manton@damtp.cam.ac.uk}} \\[20pt]

\vskip 1em
{\it 
Department of Applied Mathematics and Theoretical Physics,\\
University of Cambridge, \\
Wilberforce Road, Cambridge CB3 0WA, U.K.}
\vspace{12mm}

\abstract
{The asymptotic partition function for quantized Abelian Higgs
vortices at high temperature $T$ is found to leading and
subleading order, and from this the equation of state of the vortex gas
is derived, including the first quantum correction. It is
assumed that the Hamiltonian is proportional to the Laplace--Beltrami
operator on the moduli space of static $N$-vortex solutions. The partition
function is calculated using the total volume and total scalar
curvature of the moduli space.}

\end{center}

\vskip 150pt
\leftline{Keywords: Abelian Higgs vortices, Quantum statistical
mechanics, Moduli space}
\vskip 1em

\vfill
\newpage
\setcounter{page}{1}
\renewcommand{\thefootnote}{\arabic{footnote}}


\section{Introduction} 
\vspace{4mm}

In \cite{Man}, the author studied the classical statistical mechanics of
Abelian Higgs vortices at critical coupling (i.e., BPS vortices), where the
static forces between vortices cancel. It is assumed that there are $N$
vortices at temperature $T$ on a large spherical surface of area $A$, where
$A > 4\pi N$. The last inequality, due to Bradlow \cite{Bra}, ensures
that static vortex solutions exist, and its interpretion is that each
vortex occupies an area $4\pi$, although a vortex is not a hard disc. The
$N$-vortex classical dynamics is then assumed to be modelled by geodesic
motion through the moduli space $\M$ of static $N$-vortex
solutions \cite{Ma1}. The vortex number $N$ is conserved, because it is a
topological invariant. One is interested in the statistical mechanical
equilibrium state in the thermodynamic limit, where $N,A \to \infty$
and the number density $N/A$ remains finite. 

In this model of a vortex gas, the partition function reduces to a simple
Gaussian integral over momenta, dependent on the temperature $T$,
multiplied by the volume of the moduli space $\M$. Fortunately, the volume
can be calculated exactly. $\M$ has a K\"ahler metric,
for which a localised expression was discovered by Samols \cite{Sam}.
Although this metric is not known explicitly, it is possible to find
exactly the real cohomology class of the K\"ahler form, and from this
the volume can be determined by simple algebra. The coefficient
of this class (relative to a basis element for the integer cohomology)
is related to the effective mass of a cluster of $N$ coincident vortices,
moving coherently around the sphere. This effective mass can be
calculated using the Samols metric and symmetry arguments, and is
less than $N$ times the mass of one vortex.

From the partition function, it is straightforward to find the
pressure $P$ and hence the equation of state of the vortex gas,
using standard thermodynamic formulae \cite{LL}. The result is
\be
P = \frac{NT}{A-4\pi N} \,.
\label{Clausius}
\ee
This is the Clausius equation of state for a 2-dimensional non-ideal
gas, where the gas particles each take up some area that is excluded to
others, but otherwise exert no static forces on each other.

Remarkably, this is an exact equation of state, given our assumptions,
illustrating that the moduli space dynamics of Abelian Higgs vortices is an
elegant topic within mathematical physics. In the partition function,
the indistinguishability of the vortices is inescapable, so a Gibbs
factor of $N!$ appears automatically and does not need to be
introduced by hand. Hard discs have something in common with
vortices, but the equation of state for a gas of hard discs is
not known exactly, and beyond the second virial coefficient, it
differs from eq.({\ref{Clausius}}). On the other hand, eq.({\ref{Clausius}})
is very similar to the equation of state for a gas of hard rods of
finite length, moving along a line \cite{Ton}. This similarity has been
explained through the use of T-duality by Eto {\it et al.} \cite{Eto}.

One might worry that the above equation of state is sensitive
to the curvature of the background spherical surface, even though this
curvature is very small in the thermodynamic limit. This was investigated by
Nasir and the present author \cite{MN}, who found the volume of the $N$-vortex
moduli space $\M$ for vortices on a compact surface of any genus
$g$ and area $A$, having any smooth Riemannian metric. As a
manifold, $\M$ is the symmetrised $N$th power of the underlying
surface. Again, the Bradlow inequality $A - 4\pi N > 0$ needs to
be satisfied for $N$-vortex solutions to exist.

The metric on $\M$, in this general situation, is a
variant of the Samols metric, and is again K\"ahler. The K\"ahler
2-form for $g \ge 1$ is a real combination of two basic, integer
cocycles, whose coefficients can be calculated by considering the
2-cycle where $N$ coincident vortices move together, and the 2-cycle where
one vortex moves and the remaining $N-1$ are coincident and fixed.
These coefficients have been confirmed by Perutz, using an independent
argument \cite{Per}. The total volume of $\M$ is then calculated from
the $N$th power of the K\"ahler 2-form, divided by $N!$, using the
cocycle algebra established by Macdonald for symmetrised powers of
Riemann surfaces \cite{Mac}. This calculation gives an explicit formula
for the volume, depending only on $g$, $N$ and $A$. Although the
volume is genus-dependent, one finds that this dependence drops out in
the thermodynamic limit, and the equation of state is independent
of genus.    

In this paper, we will recalculate the equation of state for Abelian
Higgs vortices using quantum statistical mechanics. The quantum
Hamiltonian will be taken to be $\half\hbar^2$ times the
Laplace--Beltrami operator on the moduli space $\M$, the quantized version
of the classical kinetic energy for the vortex gas. We will need to assume
that the temperature $T$ is large and that $\hbar$ is small, in senses
to be clarified below. We can then use the asymptotic, large $T$
expansion for the partition function \cite{BGM}. The leading term
depends on the volume of the moduli space $\M$, and reproduces the
classical partition function. The subleading term depends on the
total scalar curvature of $\M$, and is proportional to $\hbar^2$.
This total curvature is again calculable, because
$\M$ is a K\"ahler manifold, and it has been found by Baptista
\cite{Bap}. The calculation combines the real cohomology class of
the K\"ahler 2-form on $\M$ with the first Chern class of (the tangent
bundle of) $\M$.  Both of these have known coefficients in terms of
basic, integer cocycles, so the Macdonald algebra can again be used to
complete the calculation. The total volume and total scalar curvature
of the $N$-vortex moduli space for general genus $g$ are found to be sums
of $g+1$ terms, but these sums simplify in the thermodynamic limit. The
subleading curvature term gives the novel result of this
paper, the first quantum correction to the classical partition
function at high temperature, and hence the first quantum correction
to the equation of state. In particular, we find the quantum
correction to the second virial coefficient of the vortex gas.

\vspace{4mm}

\section{$N$-vortex partition function}
\vspace{3mm}

We will not review here the fundamental, first order field equations for
static vortex solutions in the Abelian Higgs model at critical
coupling. For this, see \cite{Tau,book}. We just recall that these
equations can be defined on any compact Riemann surface $\Sigma$ of
genus $g$, equipped with an arbitrary, smooth Riemannian metric
compatible with the complex structure (i.e., having the given
conformal structure). $N$-vortex solutions exist provided $A > 4\pi N$,
where $A$ is the area of $\Sigma$. Up to gauge transformations, there
is a unique solution for any set of $N$ (not necessarily distinct)
unordered points on $\Sigma$ -- the points where the Higgs field
vanishes \cite{Bra,Gar}. These solutions minimise the field potential
energy in the $N$-vortex sector, and they all have the same energy, a
constant multiple of $N$. The moduli space of solutions, $\M$, is
therefore the $N$th symmetrised power of $\Sigma$. $\M$ inherits the
complex structure of $\Sigma$, and is a smooth manifold with complex
dimension $N$.

One may assume that the dynamics of $N$ vortices with small kinetic
energy is geodesic motion through the moduli space $\M$ \cite{Ma1}. The
field kinetic energy is defined as a gauge-invariant integral over $\Sigma$,
quadratic in the time-derivatives of the fields, and when restricted
to motion tangent to $\M$, it defines the Riemannian metric on
$\M$, which is actually K\"ahler. Using the linearised field equations
and Gauss's law, Samols reduced this integral to a compact, elegant
formula for the metric, depending only on local information about
the Higgs field close to each vortex \cite{Sam}. The
Samols metric on $\M$ is smooth. Physically, this means that vortices
scatter smoothly even if they are instantaneously coincident. This is
because the underlying field dynamics is smooth, and vortices have no
point singularities.

Let $\{q^i : 1 \le i \le 2N \}$ denote real coordinates on
$\M$, and $g_{ij}({\bf q})$ the Samols metric. Then the kinetic energy 
for motion through moduli space is simply
\be
L = \half g_{ij}({\bf q}) {\dot q}^i {\dot q}^j \,.
\ee
This is the complete Lagrangian, if we drop the constant potential energy,
and so it is the complete classical energy. [Note: Samols
\cite{Sam} has an additional explicit factor of $\pi$ here,
representing the mass of one vortex. We follow Baptista \cite{Bap}
in absorbing this into the metric.]
The conjugate momenta are
\be
p_i = \frac{\pr L}{\pr q^i} = g_{ij}({\bf q}) {\dot q}^j \,,
\ee
and the classical Hamiltonian, expressed in terms of the momenta, is
\be
H = \half g^{ij}({\bf q}) p_i p_j \,,
\ee
where $g^{ij}({\bf q})$ is the inverse metric.

We assume that the quantized vortex dynamics can be treated as the
quantized, purely kinetic dynamics on the moduli space $\M$. This may be an
oversimplification, because the quantum field theory may generate
a non-constant potential energy on $\M$ -- however, we ignore this
possibility. The quantized conjugate momentum operators,
\be
p_i = -i \hbar \frac{\pr}{\pr q^i} \,,
\ee
are inserted into the classical Hamiltonian to determine the quantum
Hamiltonian. There is a familiar operator-ordering problem here,
resolved in the standard way by assuming that the final quantum
Hamiltonian is
\be
H = \half \hbar^2 \Delta \,,
\ee
where $\Delta \equiv -\nabla^2$ is the Laplace--Beltrami operator on $\M$.
Since $\M$ is compact, without boundary, the spectrum of $\Delta$ is a
discrete set of non-negative eigenvalues $\lambda$. The partition
function at temperature $T$ is then
\be
Z(T) = \sum_{\lambda} e^{-\frac{\hbar^2}{2T} \lambda} \,.
\label{partition}
\ee

Let us now clarify the order of magnitude of the parameters here. In
the thermodynamic limit, $N$ is large, so the Bradlow inequality
requires $A$ also to be large. The field theory
Lagrangian of the Abelian Higgs model has $O(1)$ coefficients, so the
vortex mass is $O(1)$. Because of the Higgs mechanism, there is a
massive photon and a massive scalar particle in the quantized field
theory. Their physical masses are $O(\hbar)$. We assume $\hbar \ll 1$,
so that we are in the perturbative regime of the $N$-vortex sector of
the theory, where vortices
are heavy relative to the other particles. We need the temperature
$T$ to be small enough, so that photons and scalar particles are
not created in vortex collisions. Therefore $T \ll \hbar$.
Additionally, we need the $N$-vortex dynamics to be close to the
classical regime, so $T$ must not be too small. The typical, small
energy eigenvalue for a quantized single vortex moving on a surface of area
$A$ is $O(\hbar^2/A)$, so we require $T \gg \hbar^2/A$. These
conditions can be satisfied if $T$ is $O(\hbar^2)$.

We cannot calculate the partition function (\ref{partition}) exactly, but
it has a well-known asymptotic expansion in $\hbar^2/2T$ -- this is
essentially the asymptotic expansion of the trace of the heat kernel
for the Laplace--Beltrami operator on $\M$. As $\M$ has real
dimension $2N$, the first two terms of this expansion are \cite{BGM}
\be
Z(T) = \left( \frac{2\pi\hbar^2}{T} \right)^{-N} \left( {\rm Vol}(\M)
  + \frac{\hbar^2}{12 T} \, {\rm Curv}(\M) \right) \,,
\label{twotermpart}
\ee
where
\be
{\rm Vol}(\M) = \int_{\M} d{\rm vol}_{\M}
\ee
is the $2N$-dimensional, total volume of $\M$ and
\be
{\rm Curv}(\M) = \int_{\M} s \, d{\rm vol}_{\M}
\ee
is the total curvature, the integral over $\M$ of the scalar
curvature $s$, where all the geometrical quantities are determined
from the Samols metric.

Baptista has calculated the total volume and total curvature of the
moduli space for a number of types of vortex -- see Theorem 5.1 in
ref.\cite{Bap}. We only need the results for the Abelian Higgs
model, and to match conventions and notation, we set $\tau = 1$, $e^2
= \half$, $n=1$, $d=N$ and ${\rm Vol} \, \Sigma = A$ in Baptista's
formulae. Baptista's integer $r$ equals $N$.

For general genus $g$, the total volume is    
\be
{\rm Vol}(\M) = \pi^N \sum_{i=0}^g \frac{g!}{i! (N-i)! (g-i)!}
(4\pi)^i (A - 4\pi N)^{N-i} \,,
\ee
in agreement with ref.\cite{MN}. In the thermodynamic limit, with $g$
fixed and $N \gg g$, we can replace $(N-i)!$ by $N!/N^i$. Then
\be
{\rm Vol}(\M) = \frac{\pi^N}{N!} (A - 4\pi N)^N
\sum_{i=0}^g \frac{g!}{i!(g-i)!}\left( \frac{4\pi N}{A - 4\pi N}
\right)^i
\ee
which is just a finite binomial series, so
\bea
{\rm Vol}(\M) &=& \frac{\pi^N}{N!} (A - 4\pi N)^N \left(1 + \frac{4\pi
    N}{A - 4\pi N} \right)^g \nonumber \\
&=& \frac{\pi^N}{N!} (A - 4\pi N)^{N-g} A^g \,.
\eea
The total curvature is \cite{Bap}
\be
{\rm Curv}(\M) = 2 \pi^N \sum_{i=0}^g \frac{g!(N+1-2g+i)}{i!(N-1-i)!(g-i)!}
(4\pi)^i (A - 4\pi N)^{N-1-i} \,.
\ee
In the thermodynamic limit, we can replace $N+1-2g+i$ by $N$ and
$(N-1-i)!$ by $N!/N^{1+i}$. The sum again simplifies to a binomial
series, and the result is
\be
{\rm Curv}(\M) = \frac{2N^2\pi^N}{N!} (A - 4\pi N)^{N-g-1} A^g \,.
\label{Curvg}
\ee

The two-term partition function (\ref{twotermpart}) is therefore
\be
Z(T) = \left(\frac{T}{2\hbar^2}\right)^N \frac{1}{N!} (A - 4\pi N)^{N-g}
A^g \left[ 1 + \frac{\hbar^2 N^2}{6T(A - 4\pi N)} \right] \,.
\label{twotermresult}
\ee
The first term in the square bracket gives the partition function of
classical statistical mechanics, agreeing with refs.\cite{Man,MN}, and the
second, curvature term is the leading quantum correction.

It is worthwhile to understand heuristically the magnitude of the
total scalar curvature. Roughly speaking, it has two contributions.
Firstly, there is the generic region of the moduli
space where all vortices are well separated. For a single vortex on
a surface $\Sigma$ of area $A$, the local curvature is $O(1/A)$, so the
scalar curvature for $N$ vortices is $O(N/A)$, as this involves a
trace. The integration region has volume $O(A^N/N!)$, so the total
curvature of the generic region is $O(A^{N-1}/(N-1)!)$. This part can
be positive, negative or zero. Secondly, there is the region where one
pair of vortices is close together, and the remaining $N-2$ vortices
are well separated. The curvature of the moduli space for a vortex pair,
when they are close together, is $O(1)$ and positive, as shown by
Samols \cite{Sam}, whereas the remaining vortices contribute curvature
$O((N-2)/A)$. At low density, the $O(1)$ pair curvature dominates. The
volume of the region of the moduli space where one pair is close
together is $O(A^{N-1}/(N-2)!)$, because the pair acts like a double
vortex that is distinct from the remaining $N-2$ vortices. The total
curvature from this region is therefore $O(A^{N-1}/(N-2)!)$ and
positive. This dominates, by a factor of order $N$, the contribution
from the generic region where all vortices are well separated, and is
the order of magnitude for the total scalar curvature, agreeing with
the result (\ref{Curvg}) to leading order in $N/A$.

\vspace{4mm}

\section{Free energy and equation of state}
\vspace{3mm}

We now use the trick, discussed by Landau and Lifshitz \cite{LL}
(Sect. 72), of assuming temporarily that the density $N/A$ is small and also
that the total number of vortices $N$ is relatively small, so that
$N^2/A$ is small. The effects of two-vortex interactions are still
correctly accounted for, so extensive thermodynamic quantities and
the equation of state can be scaled up to large $N$ at the end.

The free energy of the vortex gas is $F = -T\log Z$, where $Z$ is
given by the expression (\ref{twotermresult}). 
We can make the approximation $\log(1+x) \simeq x$ for the terms in
square brackets, because of the above-mentioned trick, and we also
use the Stirling approximation $\log N! = N\log N - N$.
The free energy then becomes
\bea
F &=& -T\Biggl\{ N\log \left( \frac{T}{2\hbar^2}\right) - N\log N + N
  + (N-g)\log(A - 4\pi N) \nonumber \\
  && \qquad + g\log A + \frac{\hbar^2 N^2}{6T(A - 4\pi N)} \Biggr\} \,.
\eea
Compared with the extensive terms proportional to $N$, the
$g\log A$ term is negligible, and can be replaced by $g\log(A - 4\pi N)$.
$F$ then has the form
\bea
F &=& -T\Biggl\{ N\log \left( \frac{T}{2\hbar^2}\right) - N\log N + N
  + N\log(A - 4\pi N) \nonumber \\
  && \qquad + \frac{\hbar^2 N^2}{6T(A - 4\pi N)} \Biggr\} \,,
  \eea
which is independent of the genus $g$.  

We are mainly interested in the pressure of the gas. This is
\be
P = -\frac{\pr F}{\pr A} = \frac{NT}{A - 4\pi N}
- \frac{\hbar^2 N^2}{6(A - 4\pi N)^2} \,,
\ee
which can be rearranged into the equation of state
\be
\left( P + \frac{\hbar^2 N^2}{6(A - 4\pi N)^2} \right)(A- 4\pi N) = NT \,.
\ee
If we drop the quantum correction proportional to $\hbar^2$,
this agrees with the Clausius equation of state found using classical
statistical mechanics. The quantum correction converts the equation of state
to the van der Waals form, if we approximate $A - 4\pi N$ by $A$ in
the correction term.

The low-density expansion for the pressure is
\be
P \simeq \frac{NT}{A} \left( 1 + \left(4\pi -
    \frac{\hbar^2}{6T}\right)\frac{N}{A} \right) \,,
\ee
so the second virial coefficient is $B(T) = 4\pi - \frac{\hbar^2}{6T}$. 
The quantum correction therefore results in a decrease of $B(T)$ from
its classical value as $T$ decreases from infinity. However, the
validity of the truncated asymptotic expansion for $Z(T)$
requires that $\frac{\hbar^2}{T} \ll 1$, so we cannot conclude
without further work that $B(T)$ changes sign at some temperature
$T$ of order $\hbar^2$. It would be interesting if such a sign change
did occur -- the interpretation would be that the effective repulsive
interaction of vortices, due to their finite area, becomes an effective
attraction at low temperature due to their bosonic nature.

\vspace{4mm}

\section{Conclusions}
\vspace{3mm}

We have calculated the first quantum correction to the free energy and
equation of state of a gas of critically-coupled Abelian Higgs
vortices -- BPS vortices -- using knowledge of the volume and total
scalar curvature of the moduli space of static $N$-vortex solutions. The
equation of state has a van der Waals form. The results are valid when the
temperature is high enough so that $N$-vortex dynamics is approximately
classical, but not so high that massive photons and scalar particles
are created. It would be interesting to extend these results to other
types of BPS vortices. This should be possible using Baptista's rather
general formulae for the moduli space volume and total scalar
curvature \cite{Bap}. It would also be interesting to determine the
thermodynamic properties of vortex gases at lower temperatures. This will
require a more precise understanding of the energy spectrum of quantum
states on the $N$-vortex moduli space.

\vspace{3mm}

\section*{Acknowledgements}

I am grateful to Jo\~ao Baptista for very helpful comments on an 
earlier version of this paper, and to an anonymous referee for
critical comments on the range of validity of the results here.

\end{document}